# FIM tips in SPM: Apex orientation and temperature considerations on atom transfer and diffusion

**William Paul, David Oliver, Yoichi Miyahara, and Peter Grütter**

Department of Physics, Faculty of Science, McGill University, Montreal, Canada.

E-mail: paulw@physics.mcgill.ca

**Abstract**
Atoms transferred to W(111) and W(110) tip apices from the Au(111) surface during tunneling and approach to mechanical contact experiments in STM are characterized in FIM at room temperature and at 158 K. The different activation energies for diffusion on the (111) and (110) tip planes and the experiment temperature are shown to be important factors controlling the extent of changes to the atomic structure of the tip. W(111) tips are much better suited to scanning probe studies which require the characterization of an atomically defined tip and subsequent verification of its integrity in FIM. The statistics of the observed spikes in the tunneling current when the tips are approached to Au(111) are interpreted using a simple model of adatoms diffusing through the STM junction.

## 1. Introduction

The use of tips defined by field ion microscopy (FIM) in scanning probe microscopy (SPM) has several notable advantages [1]: an atomically defined tip will pre-define the lateral imaging resolution, the chemical nature of the apex is guaranteed, and the electronic structure is calculable based on the known geometry. FIM tips offer great potential for the understanding of contrast mechanisms in scanning tunneling microscopy (STM) and atomic force microscopy (AFM) where the atomic configuration of the tip is expected to be of great importance [2], but is usually experimentally uncharacterizable. These tips are also well suited for atomic-scale nanoindentation where the tip geometry is needed to understand the initiation of plasticity as well as the electronic conductance of the junction [3,4]. With the recent developments of the qPlus [5] and length extension resonators [6] in AFM which often employ tungsten tips appropriate for FIM [7,8], we expect the number of instruments using atomically characterized FIM tips to grow in the near future.

The implementation of an atomically defined FIM tip in simultaneous AFM and STM opens up the possibility of performing force and current characterization of an atomically defined nanojunction, perhaps connecting to a single molecule, where the positions of all relevant atoms are controlled. In a single molecule junction, the exact atomic arrangement of the metallic contacts affects metal-molecule coupling, energy-level lineup, and the electrostatic potential profile across the junction resulting in considerable changes to I-V curves [9,10]. To rigorously test and contribute to the refinement of theoretical modeling of nanoscale structures, one needs data from experiments in which the atomic-scale contact geometry is known and controllable.

In such atomically-defined SPM experiments, tip integrity is of paramount importance – one must be able to characterize a probe apex in FIM and transfer it to the SPM experiment with certainty that the atomic arrangement at the end of the tip is unchanged. Tungsten tips are highly reactive and will readily dissociate and chemically bind with gases. Even in ultra-high vacuum conditions, one must be careful to keep impurity gases at bay especially during and after admission of the FIM imaging gas (usually helium). The transfer between imaging modes must also be relatively prompt in order to maintain statistical confidence that no rest gases have adsorbed. This has been a subject of a previous investigation, in which we developed the 'force-field' method of preserving the atomic integrity of the tip using a large electric field to ionize and repel any impinging rest gas molecules [11]. This method allowed us to controllably approach an atomically-defined FIM tip to a cleaved silicon surface in STM and demonstrate its return to FIM with an unchanged apex.

After taking care that the FIM tip does not react with impurity gases in the vacuum chamber, one must carefully approach it to the sample surface without overshoot of the feedback controller [12]. Finally, tip changes (due to tip-sample interactions, for example) must be absent from an atomically-defined experiment to be assured that the apex structure remains intact. We have noted that scanning Au(111) and HOPG surfaces at low tunneling current conditions (6 pA) at room temperature leads to completely altered FIM tips, while scanning cleaved Si(111) can be carried out for some minutes with no detectable tip alterations. The absence of tip alterations in FIM is also correlated with the absence of any tunneling current instabilities (spikes) during the experiment.

Here, we explore tip integrity and the resolvability of tip changes in FIM in greater detail by approaching tips of differing apex orientations (W(111) and W(110)) to tunneling proximity with Au(111) surfaces at temperatures of 298 K and 158 K. The tips are additionally approached to mechanical contact to induce changes to their atomic structure and are subsequently characterized by FIM. At both of these temperatures, gold is uncontrollably transferred to the tip.

In the case of the W(111) tip, the transferred atoms can diffuse readily on the tip surface at room temperature, but do not diffuse when the system is cooled to 158 K. The smooth close-packed planes of the W(110) surface, however, still allow the surface diffusion of transferred adatoms at 158 K. In contrast to W(110) tips, W(111) tips are better suited to SPM studies with atomically defined probes because their large surface corrugation hampers the diffusion of transferred atoms and also permits atomic resolution at the apex region in FIM.

The experimental results presented in this paper are separated into two main sections: Section 3 presents FIM characterization of W(111) and W(110) tips after tunneling and mechanical contact experiments with Au(111) surfaces. In section 4, the statistics of tunneling current spikes are interpreted by invoking a simple model of a diffusing adatom momentarily altering the conductance of the STM junction, and we comment on the prospects and challenges of using this method for the study of surface diffusion.

## 2. Experimental methods

Experiments were carried out in ultra-high vacuum (UHV) at room temperature and at 158 K (temperature of the tip and sample during both FIM and STM). Au(111) substrates were prepared by epitaxial growth of Au on mica to a thickness of 100 nm (these samples were rigidly anchored in order to minimize tip-sample mechanical noise – they were not mounted in the cantilevered geometry used elsewhere [3,4]). The Au(111) surfaces were cleaned by repeated 1 keV Ne$^+$ ion sputtering and annealing cycles in UHV to several cycles beyond the disappearance of carbon in Auger electron spectroscopy. A STM topograph of a clean Au(111) surface is shown in Figure 1(c)

Tips were electrochemically etched from polycrystalline tungsten or single-crystalline W(111) wire and prepared by flash annealing and degassing cycles in UHV [13,14]. Tips fabricated from polycrystalline tungsten wire nearly always terminate with a (110) oriented grain at the apex (to within several degrees) due to the crystallographic texture arising from the cold drawing process used to fabricate the wire [15,16]. Field evaporation was used to prepare a clean tip surface by raising the imaging field by ~10-20% during FIM relative to the field required for He$^+$ ion imaging of the apex. The FIM image of a 6.7 ± 0.8 nm radius W(111) tip apex is shown in Figure 1(a) with the low-index planes labeled. The radius is determined by the ring counting method [11,17–19]. The apices of the W(111) tips end in three individually resolved atoms (trimer), similar to the tips described in Ref. [11].

After preserving the atomic integrity of the tips using the 'force field' protocol while UHV conditions recovered following FIM [11], the tips were approached to tunneling interaction with Au(111) samples at a setpoint of 6 pA at −0.08 V sample bias without overshoot of the tunneling setpoint. The initial coarse approach was monitored optically: the tip can be brought to 5-10 μm from the surface by observing its reflection off the sample, allowing the sample surface to be found less than 30 minutes after removing the 'force field' voltage from the tip. Upon finding the sample surface, the tunneling current was recorded at low feedback gain for later analysis. In the case of the W(110) tips made from polycrystalline wire, soft mechanical indentations were carried out as described in the results section. FIM was then performed again on the tip apices to examine modification caused during the tunneling or mechanical contacts experiments. The FIM and STM microscopes are combined into the same unit which enables the switching of modes without any transfer of the tip.

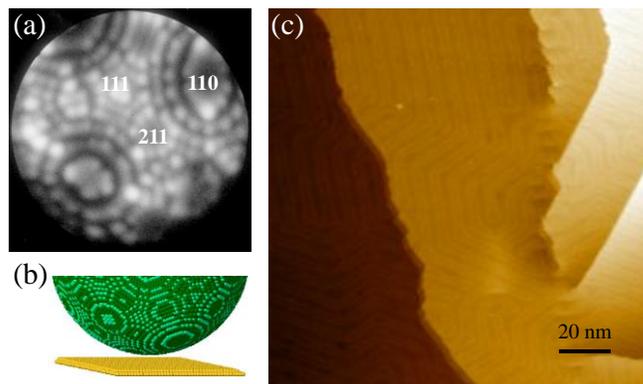

Figure 1: (a) Typical W(111) trimer tip prepared by field evaporation having a radius of 6.7 ± 0.8 nm, imaged at 7.3kV in FIM; (b) Ball model (side view) of a W(111) tip and Au(111) surface; (c) Room temperature STM image of a clean Au(111) surface showing the herringbone reconstruction (8 pA, -0.95 V sample bias).

## 3. Atom transfer to W(111) and W(110) tips

*3.1. W(111) tip / Au(111) surface at 298 K and 158 K*

When clean FIM tips are approached to tunneling proximity with Au(111) surfaces, spikes are always observed in the tunneling current. At room temperature, the spikes reach a maximum of ~40 pA, and lead to a completely changed tip structure after remaining within tunneling proximity of the sample for a few minutes, as illustrated in Figure 2(a). The same type of tunneling experiment was again performed with the Au(111) surface but at a temperature of 158 K, as shown in Figure 2(b). A representative snapshot of the current trace during tunneling and FIM images of the tip structure before and after tunneling are shown. The vertical axes of the current traces have the same limits, emphasizing that the current spikes in the lower-temperature data are much larger, reaching ~130 pA. The FIM tip retracted from the tunneling junction at 158K appears to have a nearly identical apex, with the exception of an additional atom appearing very brightly next to the original trimer apex. The lower-right atom in the trimer also appears brighter.

The minor modification of the tip apex near the (111) plane provides two encouraging results: Firstly, the minor tip changes located only at the tip apex confirm that the adatoms we observe on the FIM tips after tunneling originate from the sample, not from the tip shank. This supports our previous supposition based on the absence of tip changes when approaching the reactive cleaved Si surface [11]. Secondly, the modification demonstrates that the tip's (111) apex is indeed the part of the tip which interacts with the sample – it confirms that the apex is oriented in the correct direction. One cannot directly tell from FIM how well the crystallographic axis of the tip is aligned with respect to the macroscopic tip wire due to FIM image distortions caused by asymmetric electrostatics (caused by the tip placement with respect to the surrounding microscope design). As provided, the W(111) single crystal wire is specified to have a miscut of < 2° from the axis, and the mounting of the tip wire to the holder is also better than 2°, therefore good alignment was expected.

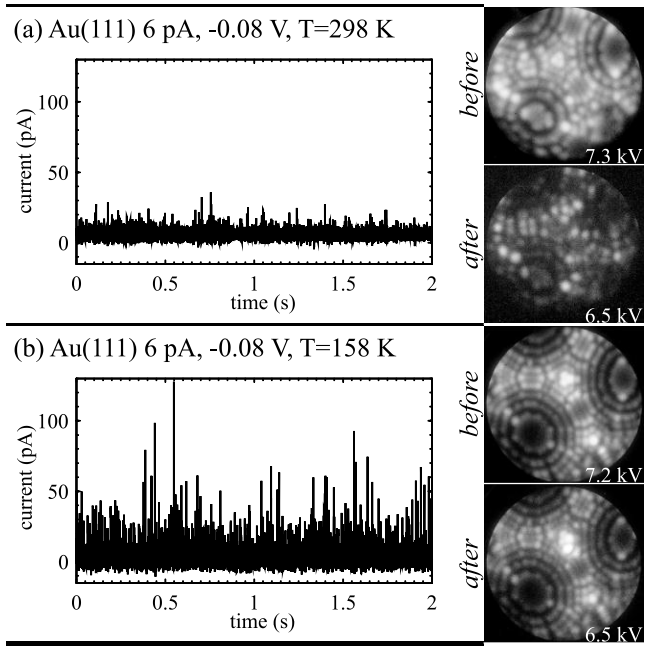

Figure 2: FIM tip apices before and after, as well as a snapshot of the measured tunneling current when approached to Au(111) at temperatures of (a) 298 K and (b) 158 K.

## 3.2. W(111) tip: field evaporation at 158 K

Several additional insights can be gained by closely monitoring the sequence of field evaporation of the tip apex when it is returned to FIM for imaging. The FIM image in Figure 2(b) was acquired slightly after the onset of imaging conditions, and corresponds to the second image of the field evaporation sequence shown in Figure 3. Each image of the sequence is an average of several photos during which the apex configuration was stable. We present a logarithmic contrast (top) and linear contrast (bottom) version of each image in the field evaporation sequence. Logarithmic contrast improves the visibility of a larger dynamic range over the entire image. Linear contrast shows the actual image contrast when viewed on the phosphorous screen, giving a more accurate representation of large differences in brightness.

The tip becomes visible in FIM at an applied voltage of about 6.1 kV, as shown in Figure 3(a). It appears from a first inspection of the logarithmic contrast image that the W(111) apex trimer is intact and an additional adatom is adsorbed just next to it. Upon increasing the voltage to 6.5 kV (Figure 3(b)), one of the trimer atoms appears much brighter – an intensity comparable to its adsorbate neighbour. At 7.2 kV (Figure 3(c)), a nearby atom to the left of the trimer begins to be imaged very brightly – the extreme contrast of these bright atoms compared to the intensity of the rest of the tip can be appreciated best in the linear contrast image. At 7.4 kV (Figure 3(d)), the first imaged adsorbate is removed by field evaporation. At 7.5 kV and up to 8.0 kV (Figure 3(e-f)), several more atoms are lost to field evaporation from the center of the tip apex. The rest of the atoms on the tip began to field evaporate in a usual homogeneous manner at ~ 8.5 kV (not shown).

The highly localized field evaporation we measure on the W(111) tips at 158 K differs substantially from the standard homogeneous field evaporation of clean W(111) tips. The localized field evaporation occurs at a comparatively low tip voltage, indicating that the adsorbed atoms promote evaporation at lower fields as well as the removal of the underlying W atoms (a common theme of adsorbed atoms on W tips [20–24]). It is not known whether Au and W evaporate as a unit as N-W and O-W have been shown to do in time-of-flight mass spectrometry measurements [23].

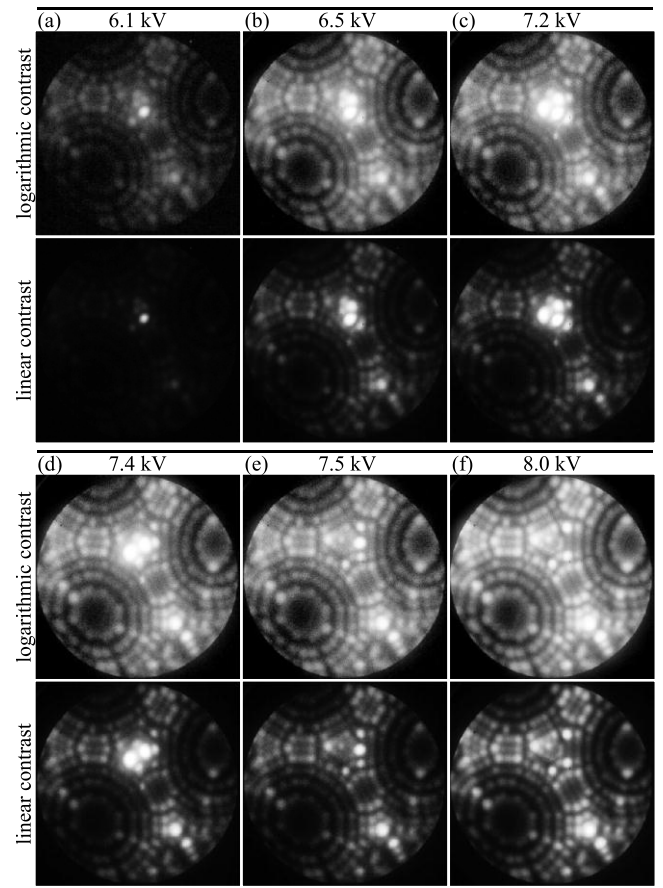

Figure 3: Field evaporation sequence of the W(111) tip apex after tunneling to the Au(111) substrate for several minutes at 158 K. Upper images have a logarithmic contrast applied to better view the full dynamic range of the FIM image. Lower images show a linear contrast to highlight the large brightness enhancement of the evaporating adatoms in FIM.

Field-induced rearrangements of the tip apex atoms certainly occur during FIM imaging, evidenced by select atoms suddenly appearing brighter before their removal by field evaporation. One might imagine mechanisms involving place exchange of W and Au atoms producing local protrusions which enhance brightness before one or several of the atoms involved in the bright spot are field evaporated. The existing theoretical literature on field evaporation [25] and equilibrium tip structures in FIM [26] is mainly concerned with homogeneous materials, not tips with small amounts of adsorbed dissimilar metals. The inspection of FIM images alone cannot reveal all the details about the transferred material or the field evaporation processes.

From this sequence of images, one can conclude that there are at least several adsorbed Au atoms on the tip – inspecting only the first image, Figure 3(a), might lead to the conclusion that only one Au atom was transferred to the tip. Apparently some of the transferred atoms can be hidden around the tip apex, perhaps at atomic plane edges just beside the apex, or in the very open and corrugated (111) surface of the bcc W crystal. A carefully recorded sequence of field evaporation images is therefore a useful experimental tool to verify the integrity of the tip apex: If indeed no tip changes occur during STM, no adsorbate promoted field evaporation would occur, and the field evaporation of the (111) apex would commence at a comparable field to the rest of the tip structure (~8.5 kV in this case).

Another concern regarding the estimation of the number of transferred atoms is that some fraction of adsorbed atoms will likely be field evaporated at a field lower than that required for ionizing the He imaging gas in FIM. Bulk Au cannot be imaged in He$^+$ ion FIM, so if a small cluster of Au formed at the tip apex, one would expect many of the bulk-like Au atoms to be removed before the onset of imaging [27,28]. This process might only allow the imaging of the last layer of Au which would be more strongly bonded to the W tip atoms. Certainly, the best way to experimentally determine the number of transferred atoms and to investigate field evaporation mechanisms is to implement time-of-flight mass spectrometry. Although the open area ratio of the channels in microchannel plate (MCP) detectors used in FIM is only ~60%, statistics could be gathered reasonably easily on the quantity and spatial location of transferred atoms. The mechanisms of field evaporation (whether or not W and Au are evaporated together) could also be determined by time-of-flight measurements.

*3.3. W(111) tip: diffusion of transferred atoms on the apex at 158 K and 298 K*

The contrast in the extent of tip modifications after tunneling with Au(111) at room temperature and 158 K is striking. Adatoms are observed to have diffused over all visible regions of the tip at 298 K. At 158 K, however, the modifications are highly localized to the apex.

The spatial extent of the tip changes observed in FIM at these two temperatures allows the estimation of rough upper and lower bounds for the activation energy of Au atoms diffusing over the W tip surface. We note that the diffusion barriers on different planes of the same crystal are in reality very different – what we extract here is an effective energy whose estimation should be dominated by the largest of the barriers of the relevant planes of the tip (in this case, the W(111) plane).

For a particle on a two dimensional surface, the mean square diffusion distance in time $\tau$ is

$$\langle r^2 \rangle = 4tD_0 \exp\left(-\frac{E_a}{k_B T}\right), \quad (1)$$

where $D_0$ is a diffusion prefactor (takes into account the attempt frequency and jump distance), $E_a$ is the activation energy of the diffusion jump process, $k_B$ is the Boltzmann constant, and $T$ is the temperature [29,30]. For an experimental delay time of $\tau \approx 1000$ s and assuming a typical prefactor of $D_0 = 10^{-3}$ cm$^2$/s [30], we plot in Figure 4 two exponential curves expressing the expected root-mean-square (RMS) displacement at temperatures of 158 K and 298 K.

From the room temperature FIM image in Figure 2(a), we can estimate a lower bound for the RMS displacement of transferred Au adatoms – they are observed essentially 'everywhere' on the tip, indicating a RMS displacement greater than 10 nm. This provides an estimate of the upper bound of the effective activation energy for diffusion of ~0.74 eV. If the energetic barrier were higher, the adatoms could not have been displaced all over the tip. From the FIM image at 158 K, Figure 2(b), it appears that the damage to the tip extends to a lateral width of only ~2-3 nm (atoms on the W(111) plane are spaced by 0.46 nm). We estimate the lower bound for the activation energy based on a RMS displacement of 1 nm at 158 K to be ~0.46 eV. It is possible that the RMS displacement due to adatom diffusion is actually less than 1 nm: If a cluster of Au formed on the tip apex, the mechanism by which the damage spread to a width of ~2-3 nm may not be diffusion of Au on the W tip, but the rearrangement of a small Au cluster on the tip as adatoms are collected. The suggested activation energy window of 0.46-0.74 eV for gold adatom diffusion on tungsten is of the correct order for metals on tungsten surfaces (see Table 1).

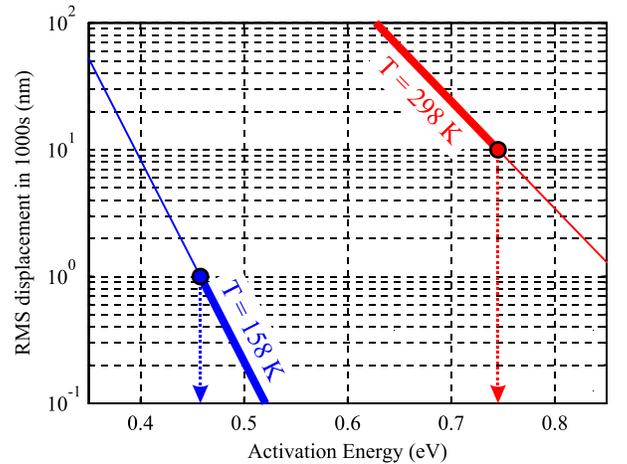

Figure 4: Expected RMS displacement of diffused adatoms at T = 158 K and T = 298 K during 1000s as a function of activation energy. Upper and lower limits of the displacement estimated from FIM data dictate a range for the activation energy for adatoms to escape from the vicinity of the tip apex.

*3.4. W(110) tip / Au(111) surface at 158 K*

In a manner similar to W(111) tips, tunneling currents measured between W(110) tips and Au(111) surfaces also exhibit continuous spikes. In FIM, the evidence of transferred atoms is much more difficult to discern than for W(111) tips because of their atomic structure made up of smooth planes. We use FIM to examine changes to the W(110) tips after approach to mechanical contact to Au(111) at 158 K and after tunneling to Au(111) at 158 K.

Figure 5 summarizes the first of two experiments in which a W(110) tip was approached to mechanical contact with the Au(111) substrate at a temperature of 158 K. The W(110) tip was fabricated with polycrystalline tungsten wire, so it is expected that the (110) pole of the apex is aligned to within 10° of the wire axis (estimated from the X-ray diffraction pole figure of the {110} reflections for cold drawn tungsten wire reported by Greiner and Kruse [16]). The tip apex was characterized by FIM before approach to the sample, shown in Figure 5(a) with the low index planes labeled.

**See end of manuscript for full-width figure**

Figure 5: (a) FIM image of the ~12 nm radius W(110) tip apex. (b) Current-distance curve acquired during approach to contact showing a large hysteresis between in and out directions. (c-f) Field evaporation sequence of the tip imaged in FIM after the approach to contact.

After the initial approach to tunneling proximity with the Au(111) surface, the tip apex was moved toward the surface to incrementally larger displacements while monitoring the current behaviour until a signature of mechanical contact was obtained. For small displacements from the tunneling setpoint, a plot of the current as a function of tip displacement, I(z), will reveal the exponential distance behaviour expected for a tunneling junction. The current recorded upon approach and retraction, 'in' and 'out' directions, should overlap unless there are any major mechanical changes occurring in the tip-sample junction. The I(z) curve plotted in Figure 5(b) shows the exponential increase on the 'in' curve (green) with an apparent barrier height of ~4.7 eV, of the order expected for a clean metal-metal contact [31]. The fit to the data is shown by the black line from 15 to 200 pA. At ~10 nA on the 'in' curve, an instability occurs and the junction conductance abruptly increases. Upon withdrawal of the tip from the sample, a large ~0.5 nm hysteresis is present, interpreted as the drawing and breaking of a wire of Au atoms pulled up from the sample. After this signature of mechanical contact, the tip was returned to FIM for characterization.

The FIM imaging sequence after mechanical contact is shown in Figure 5(c-f) during which the imaging voltage was gradually increased. In Figure 5(c), some of the changes to the tip apex are indicated by arrows – a few atoms have adsorbed onto the apex (110) plane, as well as the edge of (110) planes two to four atomic steps down from the apex. The edges of the (110) planes, particularly layers 2 and 3 from the (110) pole, also appear slightly more disordered, indicating changes to the atomic structure at their edges. Changes to the tip at the edges of (110) planes are particularly difficult to discern as the FIM does not provide resolution of the atomic structure at the plane edge. Upon increasing the imaging voltage, some adsorbed atoms begin to appear brighter (indicated with arrows in Figure 5(d-e)) before field evaporating – qualitatively similar to the behaviour of atoms seen on the W(111) tip in Figure 3.

A second experiment was performed in the same manner, summarized in Figure 6. The FIM images of the tip before and after contact are shown in Figure 6 (a) and (c), and the I(z) curve acquired during approach is plotted in Figure 6(b). Despite large thermal drift in this particular experiment making the distance axis *very* inaccurate, hysteresis and wire drawing are still indicative of a mechanical contact, especially as the conductance of the junction suddenly drops during the 'out' curve.

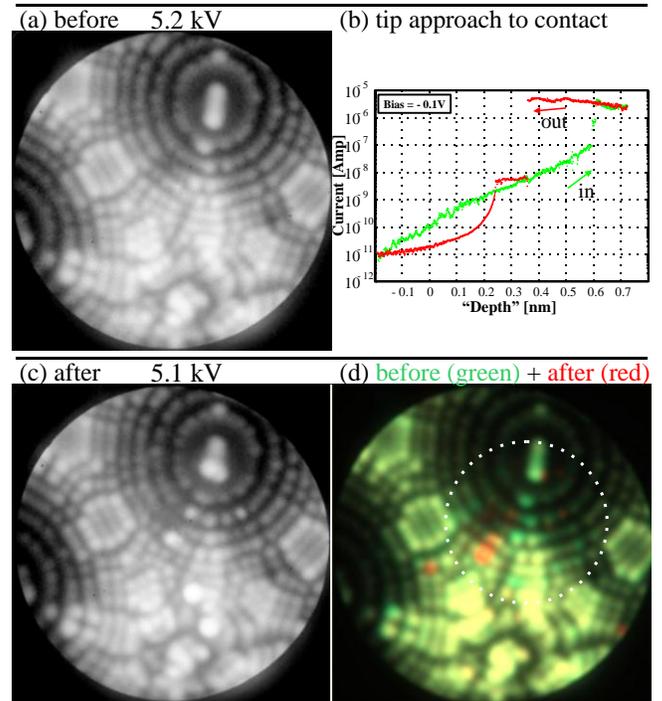

Figure 6: (a) FIM image of the ~12 nm radius W(110) tip apex. (b) Current-distance curve acquired during approach to contact. Approach ('in') and retraction ('out') directions are indicated by arrows. The distance scale is *very* inaccurate due to large thermal drift. The unusual shape of the retraction curve near its end (depth < 0.24 nm) is due to the settling time of the current preamplifier after the second break of the contact. (c) FIM image after contact experiment. (d) Colour superposition image of FIM images taken before (green) and after (red) the approach to contact.

A superposition image of the FIM images before and after contact is shown in Figure 6(d). Special attention was paid to aligning the camera so that the images could be accurately compared. The 'before' image is illuminated in green, and the 'after' image in red, so that atoms appearing in green are those that have field evaporated, and those appearing in red are newly adsorbed. The changes to the tip structure are concentrated in the region indicated by the dotted white circle, centered roughly three (110) steps down from the (110) pole.

The rather large spatial extent of the tip changes seen in these contact experiments suggest that the transferred atoms have diffused a substantial distance on the tip, in contrast with our observation for the W(111) tip, in which tip changes were concentrated to a very small region at the apex. We know that the contact area between the gold and tungsten tip is small – on the order of several atoms. The maximum conductance during the first contact experiment is ~0.05 $G_0$, and the maximum conductance in the second experiment is ~0.9 $G_0$. Ab initio transport calculations carried out to support our previous nanoindentation experiments indicated that a modified Sharvin conductance of ~0.2 $G_0$ per atom is expected due to the poor Bloch state overlap of conduction electrons in W and Au [4]. Effects of crystalline disorder could reduce this value further. In any case, the contact region is *much* smaller than the ~7 nm wide dotted circle in Figure 6(d). The diffusion rate of material away from the contact area is expected to be significantly faster on the W(110) plane compared to the W(111) plane – activation energies for single atom diffusion are generally ~2× lower than those measured for the same atomic species on the (111) plane. Activation energies obtained from FIM studies of single atom diffusion are summarized in Table 1.

Table 1: Activation energies for W, Pd, Ni adatom diffusion on W(110) and W(111) surfaces. From ref. [29].

| Surface / Adatom | W | Pd | Ni |
| --- | --- | --- | --- |
| **W(110)** | 0.93 eV | 0.51 eV | 0.49 eV |
| **W(111)** | 1.85 eV | 1.02 eV | 0.87 eV |

Although energy barriers for the diffusion of Au on W surfaces have not been measured experimentally, the scaling of the activation energies for other elements should be indicative of the behaviour expected for Au. The large atomic corrugation between binding sites on the bcc (111) surface compared to bcc (110) requires a larger energy difference between the energy minimum and saddle point – an increase of the energetic barrier for diffusion is expected regardless of adatom species. The (110) plane is the closest packed bcc crystal plane, allowing for transferred material to displace significantly, even at a temperature of 158 K.

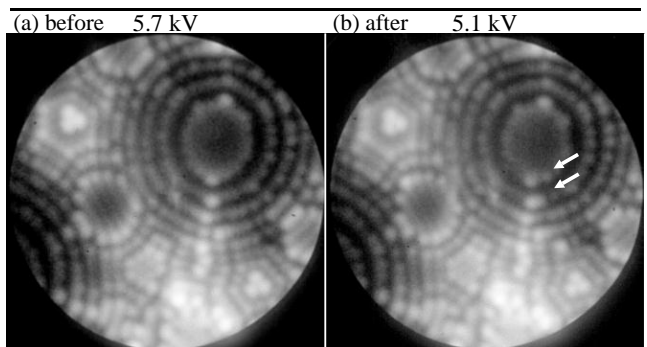

Figure 7: (a) W(110) tip apex before tunneling experiment. (b) W(110) tip apex after tunneling to Au(111) for five minutes.

W(110) tips were also approached to tunneling proximity with Au(111) substrates for several minutes to investigate the visibility of transferred atoms in a much more benign experiment than approach to mechanical contact. In these experiments, tips were approached to a setpoint of 25 pA at −0.1 V sample bias, and kept within tunneling range for five minutes. Minor changes to the (110) plane edges are seen after tunneling, shown in Figure 7(b) by arrows. Again, the (110) tip apex makes the identification of tip changes particularly difficult because of the lack of atomic resolution and the low diffusion barriers which permit the relocation of transferred atoms.

*3.5. W(110) tip – rest gas adsorption and resolution*

Finally, we comment on the resolution of adsorbed atoms in general on the W(110) planes by examining a field evaporation sequence of a tip left in UHV conditions and subjected to rest gas contamination over several days. At the beginning of the initial FIM imaging sequence, starting at 5.7 kV in Figure 8(a), a significant number of adsorbed gas atoms have already been removed, and the (111) and (211) planes have begun to appear free of bright adsorbed atoms. The (110) plane at the tip apex seems to have some brighter adsorbed atoms on its edge. From this image alone, one might conclude that the (110) plane was clean and adsorbate-free. Upon slowly increasing the field, atoms at the edge of the (110) are field evaporated from the tip, and reveal an enhanced image contrast in the central region of the plane. Figure 8(d) shows that the middle of the (110) planes is far from atomically perfect and contains many adsorbed gas atoms.

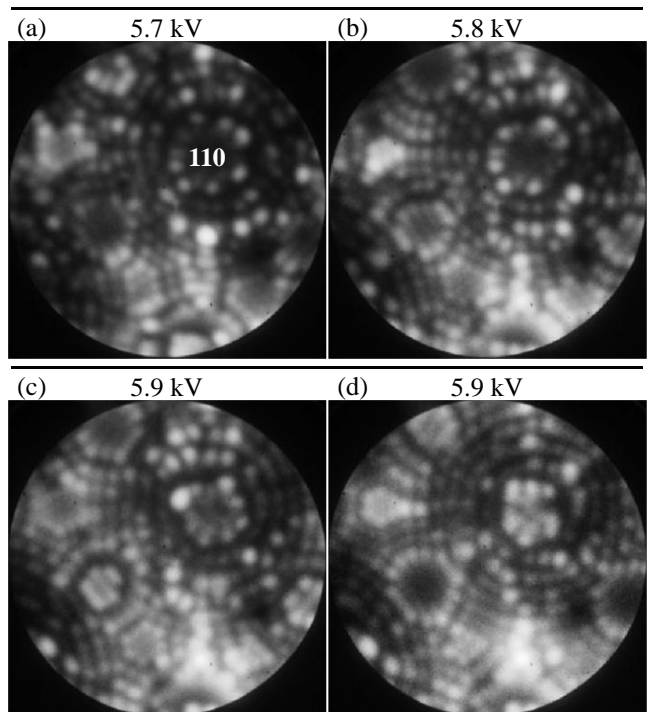

Figure 8: A field evaporation sequence of a W(110) tip left to adsorb UHV rest gases for several days. Even after other planes have been cleaned, the (110) plane is shown to hide adsorbed atoms, which begin to be imaged as the edge atoms are carefully removed.

Because the (110) planes are large and flat, the local electric field is reduced. The diminished rate of He ionization over the adsorbed atoms in the middle of the (110) plane in Figure 8(a) leads to poor contrast. Close monitoring of field evaporation is required to discern changes to the tip's atomic structure.

The large atomic corrugation of the W(111) tips makes them a more appropriate choice for atomically defined SPM studies where the identification of apex modifications is a necessity.

## 4. Tunneling current spikes

*4.1. Adatom escape model*

We now turn to the subject of the tunneling current spikes observed when clean FIM tips are approached to tunneling proximity with Au(111) surfaces. What information is contained within these spikes, and what can be learned regarding adatom motion and transfer to the tip?

The tunneling current spikes have a nearly identical peak shape, illustrated in Figure 9. Their width of ~0.4 ms is of the expected order based on the bandwidth of our tunneling current preamplifier [32]. The consistent peak shape suggests that the instantaneous current spike, due to an adatom residing momentarily in the STM junction, is very short in time. Each current spike is subjected to identical broadening from the detection electronics (the bandwidth of the feedback loop is set to be sufficiently low so that the tip-sample distance is only regulated to compensate thermal drift).

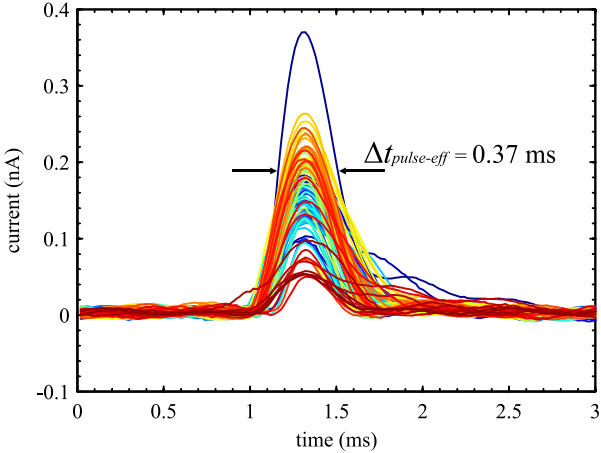

Figure 9: Tunneling current spikes detected while tunneling to a Au(111) substrate at 158 K. The peaks have a nearly identical shape suggesting that they are subjected to identical broadening due to the finite bandwidth of detection electronics.

Pictured in Figure 10(a) is the scenario of a diffusing adatom on the sample surface, depicted by a periodic potential landscape. The adatom gas density determines the rate of arrival of adatoms into the STM junction $\Gamma_{arrival}$. An adatom arriving under the tip will reside there momentarily. For a very short time, $\Delta t_{atom}$, the conductance of the tunneling junction is modified, allowing a much larger current than usual, $I_{atom}$, to pass through. The instantaneous current has the form of a rectangular pulse of width $\Delta t_{atom}$ and height $I_{atom}$, as illustrated in Figure 10(b). The pulse detected by our relatively low-bandwidth current preamplifier is smaller in height and wider in time, as shown by Figure 10(c).

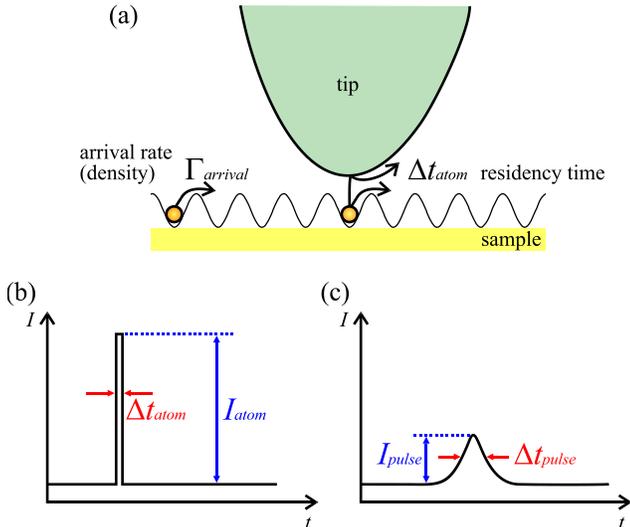

Figure 10: (a) Schematic cartoon of adatom diffusion in a periodic surface potential landscape and residing under the tip for some short time before escaping either to the tip or elsewhere on the sample. (b) The instantaneous current pulse due to the increased junction conductance is short in time and large in magnitude. (c) The detected current pulse is broadened by detection electronics, but will have the same integrated area as the pulse in (b).

The conservation of charge imposes a precise relationship between the instantaneous current spike and the broadened current pulse. The only place for electric charge to flow is through the current preamplifier, therefore the area of the instantaneous and broadened peaks must be equal:

$$Q = I_{atom}\Delta t_{atom} = \int_{t_1}^{t_2} I_{pulse}(t)dt. \quad (2)$$

Because all of the detected pulses have the same shape, we can obtain their integral by multiplying the peak height by an effective width $\Delta t_{pulse-eff}$. We find this effective width by plotting the numerically integrated peaks as a function of the maximum peak current, and taking the slope of this plot, shown in Figure 11. The obtained effective width of 0.369 ± 0.013 ms compares favourably with a visual inspection of the full-width at half-maximum of the peaks shown in Figure 9 (where we have indicated this 0.37 ms effective width to scale).

Replacing the integral in Eq. (2) with $I_{pulse}\Delta t_{pulse-eff}$, we obtain an expression for the residency time of an atom in the tunneling junction based on the pulse heights:

$$\Delta t_{atom} = \frac{I_{pulse}}{I_{atom}} \Delta t_{pulse-eff}. \quad (3)$$

In Eq. (3), a value must be estimated for the instantaneous current, $I_{atom}$. An adatom on the surface will be on the order of one lattice plane closer to the tip than the surface on which it resides, i.e. ~2.5 Å. Assuming the decay of the tunneling current to be one decade per Ångstrom, $I_{atom}$ should be on the order of $10^{2.5} \times$ greater than the tunneling current baseline in the absence of the adatom.

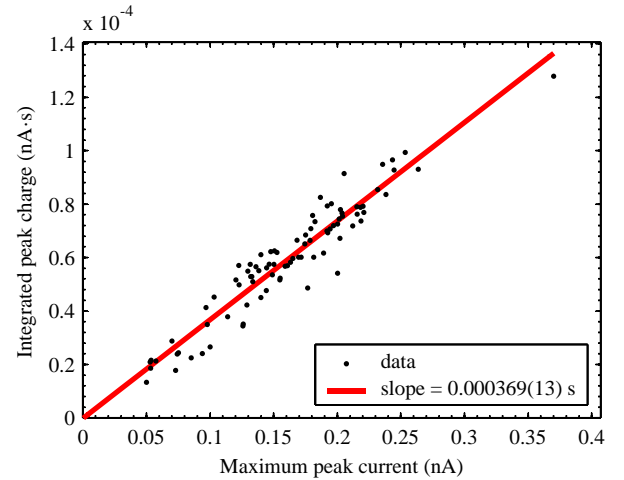

Figure 11: Plot of the integrated peak charge as a function of the maximum peak current. The slope gives the effective width which can be used to estimate the peak's area from its maximum value.

Thermal activation will lead to the adatom's escape from the tunneling junction with a rate given by the Arrhenius equation

$$\nu = \frac{1}{\langle \Delta t_{atom} \rangle} = \nu_0 \exp\left(\frac{-E_a}{k_B T}\right), \quad (4)$$

where $\nu_0$ is an attempt frequency, $E_a$ is the activation energy of the escape process, $k_B$ is the Boltzmann constant, and $T$ is the temperature. For an Arrhenius behaviour, we expect the distribution of residency times to be exponential: each attempt at frequency $\nu_0$ has an equal probability of escape, $P_{esc} =$ $\exp(-E_a/k_B T)$, thus the probability of the adatom still residing in the junction after $n$ attempts (corresponding to time $\Delta t = n/\nu_0$) is $(1 - P_{esc})^n$. A histogram of the heights of the measured current spikes is shown on a logarithmic scale in Figure 12 for the data recorded at 158 K (Figure 2(b)). The distribution of current peak heights, and thus the residency time, is exponential over at least two orders of magnitude.

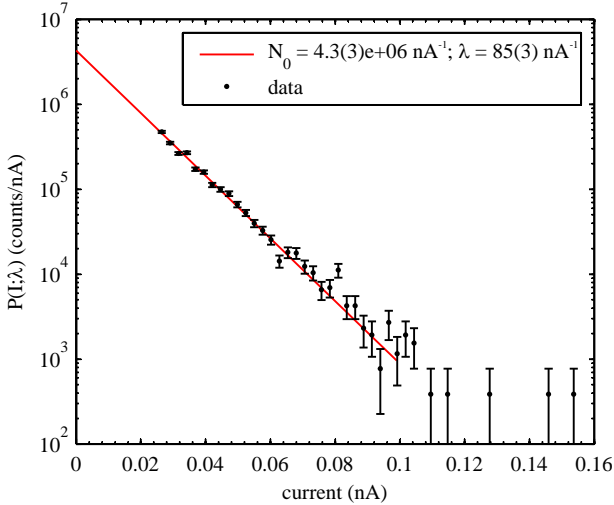

Figure 12: Log scale histogram of the number of tunneling current spikes as a function of spike height, showing an exponential distribution over at least two orders of magnitude.

By fitting the data in Figure 12 to an exponential probability distribution,

$$P(I; \lambda_I) = N_0 \exp(-\lambda_I I), \quad (5)$$

a mean pulse current of $\langle I_{pulse} \rangle = \lambda_I^{-1} = 11.8 \pm 0.4$ pA is obtained. This corresponds to a mean residency time of $\langle \Delta t_{atom} \rangle = 2.3 \pm 0.1$ μs.

Rearranging Eqs. (3) and (4), the activation energy for the escape process is given by:

$$E_a = k_B T \ln(\nu_0 \langle \Delta t_{atom} \rangle)$$
$$= k_B T \ln\left(\nu_0 \frac{\Delta t_{pulse-eff}}{I_{atom}} \langle I_{pulse} \rangle \right). \quad (6)$$

Assuming an attempt frequency of $10^{12} - 10^{13}$ Hz [30], an activation energy in the range of 0.20 – 0.23 eV is determined.

Finally, we consider whether any information about the arrival rate of adatoms, $\Gamma_{arrival}$, can be obtained from the tunneling current data. Given that the mean pulse current is about 10 pA, many of the spikes will pass below our detection threshold – simply measuring the time between the pulses detected above a threshold of ~25 pA will greatly overestimate the mean time between events. However, the total number of events in the exponential probability distribution of current pulses can be obtained by the fit parameters of the distribution (obtained by normalizing and analytically integrating Eq. (5)):

$$N_{tot} = \frac{N_0}{\lambda}. \quad (7)$$

The rate of the current spikes can be calculated by dividing this total number by the acquisition time interval. A rate of $\Gamma = 300 \pm 20$ Hz is obtained for the data in Figure 2(b) (though we believe these rates are artificially induced by tip-sample vibrations, discussed in the next section).

Performing a similar analysis on the room-temperature data, examination of the tunneling current spikes presented in Figure 2(a) yields a mean pulse current of $3.6 \pm 0.2$ pA, corresponding to a $0.70 \pm 0.04$ μs residency time below the tip. The activation energy of the escape process is estimated to be 0.35 – 0.40 eV assuming an attempt frequency of $10^{12} - 10^{13}$ Hz. The mean pulse rate is found to be $1040 \pm 300$ Hz.

*4.2. Discussion*

From the exponential distributions of tunneling current spikes obtained in several experiments at 158 K and 298 K, we obtain activation energies of $E_{a\,158\,K} = 0.19 \pm 0.01$ eV and $E_{a\,298\,K} = 0.360 \pm 0.014$ eV. The uncertainties quoted represent the standard deviation of the six measurements at 158 K and four measurements at 298 K. Here, we have taken the attempt frequency to be $10^{12}$ Hz. The measurement of two different activation barriers is somewhat unexpected for the similar experimental configurations – these may originate from two different escape mechanisms with differing energetics (for example, one for diffusion to another lattice site, and one for transfer to the tip). A process requiring 0.36 eV will have a $\sim 3 \times 10^5$ smaller rate at 158 K than at room temperature, leading to very small statistics compared to the 0.19 eV process. Meanwhile, the detection of a 0.19 eV process at room temperature would require the measurement of ~0.1 pA current spikes, which is well below our current detection noise limit.

Order-of-magnitude variations in our assumptions for the instantaneous peak current, $I_{atom}$, and the attempt frequency, $\nu_0$, contribute to errors in determined activation energies of $k_B T \ln(10)$. This translates to 0.03 eV at 158 K, and 0.06 eV at room temperature. If temperature could be varied slightly in these experiments, an Arrhenius analysis could be carried out by plotting the measured mean residency time as a function of 1/Temperature. Not only would this alleviate the required assumption of the attempt frequency, but it would also make the obtained $E_a$ independent of the estimation of the instantaneous tunneling current value, $I_{atom}$ (Eq. (3)). If the mean time were wrong by some factor $\alpha$ because of the instantaneous current estimation, it would appear at the intercept of the Arrhenius analysis, not in the slope:

$$\ln\left(\frac{1}{\langle \Delta t_{atom} \rangle}\right) = \ln(\nu_0) + \ln\left(\frac{1}{\alpha}\right) - E_a\left(\frac{1}{k_B T}\right). \quad (8)$$

The mechanism causing the tunneling current spikes has not been determined here – it is assumed that the energy barriers we measure are for the adatom to escape from the tunneling junction, which could happen by diffusion to other lattice sites on the surface, or by transfer to the tip. Certainly, the presence of spikes in the tunneling current is correlated with tip changes in FIM, as discussed earlier and in Ref. [11]. Spikes in the tunneling current could also result from rearrangements or adatoms diffusing on the tip itself. Comparison of the expected rate of adatom arrival (based on an estimated adatom gas density) to the measured rate of spikes may help discern the source of these events, and may help to illuminate the source of noise spikes in STM imaging.

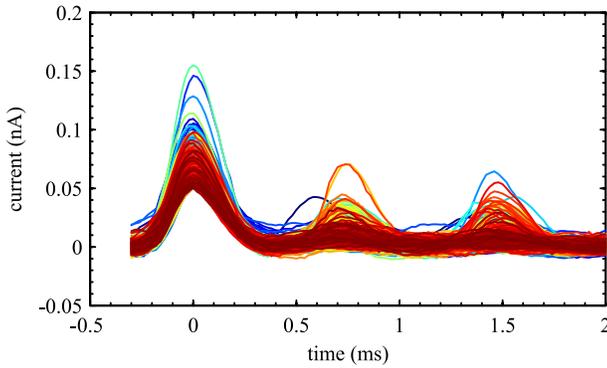
Figure 13: Time trace of the current for all spikes > 50 pA showing a regular time interval between many of the sequential spikes.

It turns out that the arrival rate of adatoms deduced earlier from the integration of the exponential curves (Eq. (7)) does not represent the true value for the unperturbed surface because of the influence of the STM tip. It is apparent that in several of our measurements, the tunneling current spikes are correlated in time. Figure 13 shows a plot of all detected peaks above 50 pA in the data collected at 158 K, centered together at time 0 ms. A periodic trend is immediately seen by eye – spikes are often followed by other spikes at regular time intervals of ~0.75 ms. The reason for this could be a very small mechanical noise which serves to modulate the tip-sample separation very slightly, resulting in a regular tilting of the adatom energy landscape. Measurements of tip-sample vibrations (by monitoring the tunneling current in the lack of distance feedback on stable Si(111)-2x1 and Cu(100) surfaces) suggest that the tip-sample gap stability is better than 20 pm. In atomistic simulations, changes in tip-sample-separation of ~30 pm have been shown to decrease barriers for diffusion or atom transfer by up to ~0.2 eV [33]. Although the arrival of spikes is correlated in time, we expect that the escape process should be unaffected by this correlation due to its relatively instantaneous timescale.

The extraction of the energy barrier for intrinsic adatom diffusion and the density of the adatom gas requires a non-interacting tip. From the direct observation of material transfer in FIM, we know that there must be a reasonable force exerted by the tip on adatoms which lowers the energetic barrier for transfer sufficiently that a finite number are transferred to the tip during measurements. Experiments performed at several tip-sample distances and applied voltages should clarify the role of the tip in the statistics of tunneling current spikes. For the moment, we have focused on reasonable low-current tunneling conditions with the main goal of assessing FIM tip integrity in STM experiments.

Modifications of the adatom potential landscape due to the presence of the tip are expected in two ways: The strong electric field in the tip-sample junction will interact with the adsorbate dipole moment, leading to a broad potential well superposed on the atomic corrugation [34,35]. In addition, the height of the energy barriers between binding sites on the surface are expected to decrease with the proximity of the tip [33].

In a more sophisticated model of adatoms momentarily residing in the STM junction, one may have to consider effects due to the resolution of the tip – an instantaneously high current might be measured while the adatom visits several sites under the tip, not just one. The known geometry of our FIM tips provides a good starting point for such estimation. If one were interested in extracting the adatom gas density, one would also have to consider the return of the same adatom into the junction due to long-range tip interactions.

## 5. Summary and conclusion

To summarize, we have shown that the transfer of atoms occurs from Au(111) substrates to W tips. Transferred material diffuses to a great extent on W(111) tips at room temperature, but is confined to the (111) apex when cooled to 158 K. W(110) tips were brought into mechanical contact with Au(111) at 158 K, and it proved to be more difficult to identify adsorbed atoms on these tips due to the larger extent of diffusion on the (110) plane. Large, flat (110) planes also hinder the characterization of the tip's atomic geometry due to the lack of atomic resolution and the poor resolution of adsorbed material within the planes and at their edges. W(111) tips provide both high diffusion barriers and atomic resolution in FIM, making them most appropriate for SPM studies with atomically defined tips. The transfer of atoms from Au(111) samples to W tips cannot be eliminated at temperatures accessible to our system.

An exponential distribution of tunneling current peak heights was observed when tunneling to Au(111) substrates at 298 K and 158 K. The mean peak heights were related to the residency time of an adatom in the STM junction through the conservation of charge and an estimation of the instantaneous tunnel current. From the residency time, the activation energy for adatom escape can be estimated. We have shown that even if not all events are above the detection noise, one can in principle integrate the exponential distribution and infer the mean rate of spikes. The influence of the STM tip, the detailed mechanisms of adatom escape, and the refinement of approximations remain open questions for future work in both experiments and atomistic modeling using the FIM tip structure as a well-defined starting point.

We conclude from this work that the tip apex orientation which defines the relevant activation barriers for diffusion, and the temperature at which experiments are carried out is of central importance to achieving a truly atomically defined experiment in SPM. In the absence of the time-of-flight characterization of the transferred material to the tip, careful examination of field evaporation sequence provides evidence of tip changes through adsorbate-promoted field evaporation. Finally, W(111) tip apices are far better suited to atomically-defined SPM experiments due to their enhanced resolution in FIM and their larger diffusion barriers compared to W(110) tips.

## Acknowledgments

Funding from NSERC, CIFAR, and RQMP is gratefully acknowledged.

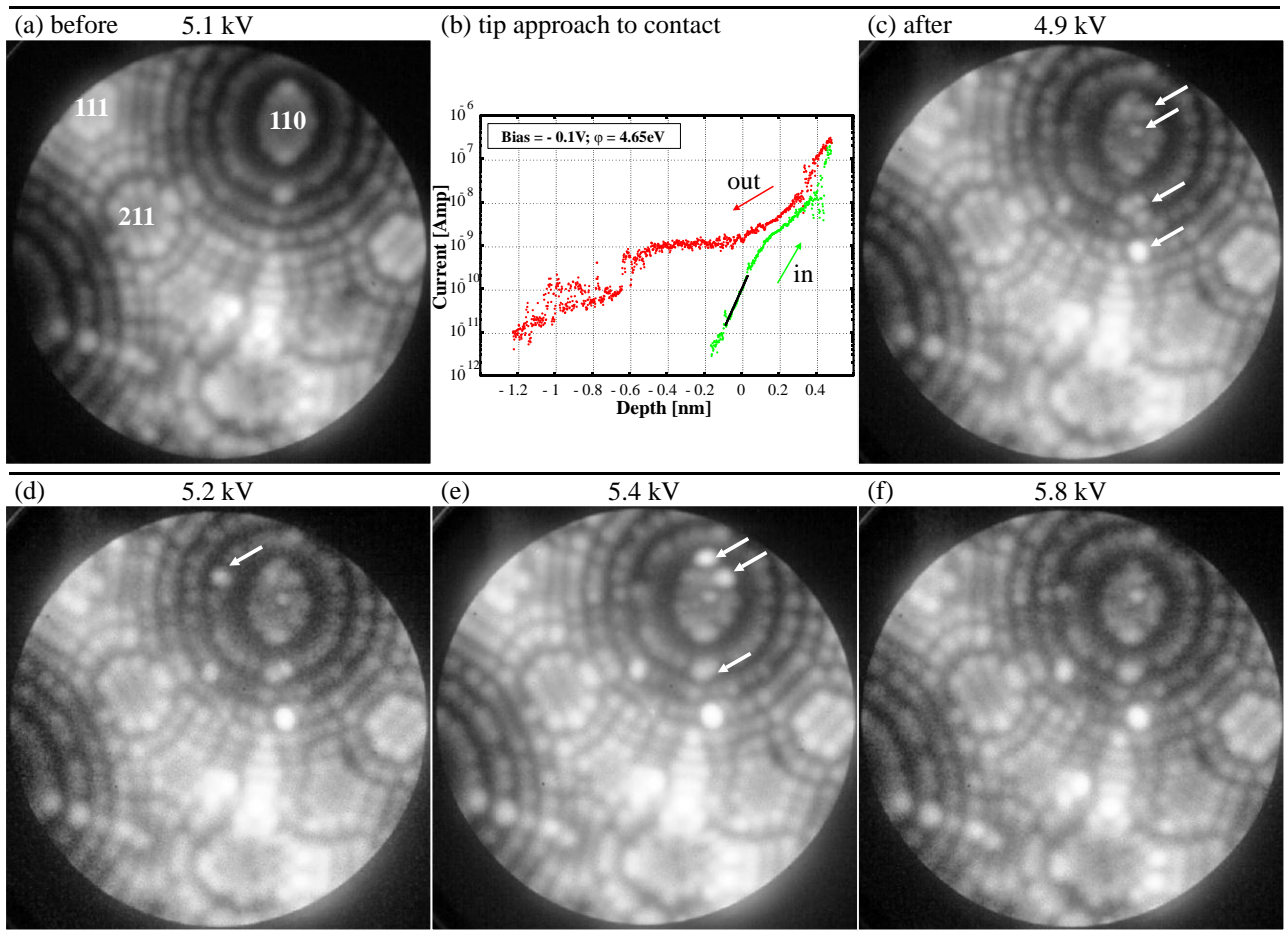

Figure 5: (a) FIM image of the ~12 nm radius W(110) tip apex. (b) Current-distance curve acquired during approach to contact showing a large hysteresis between in and out directions. Approach ('in') and retraction ('out') directions are indicated by arrows. (c-f) Field evaporation sequence of the tip imaged in FIM after the approach to contact.

11